\documentclass[reprint, superscriptaddress, secnumarabic, amssymb, nobibnotes, aps, prl]{revtex4-1}

\usepackage{graphicx}
\usepackage{epstopdf}
\usepackage[T1]{fontenc}
\usepackage{amsbsy}
\usepackage{gensymb}
\usepackage[T1]{fontenc}
\usepackage{amsmath}
\usepackage{amssymb}
\usepackage{bbm}
\usepackage{braket}
\usepackage{xcolor}
\allowdisplaybreaks
\usepackage{graphicx}
\usepackage[colorlinks=true]{hyperref}  
\hypersetup{
    bookmarks=true,         
    unicode=false,          
    pdftoolbar=true,        
    pdfmenubar=true,        
    pdffitwindow=false,     
    pdfstartview={FitH},    
    pdftitle={Superconductivity in topological Weyl Semimetal T$_d$ - Mo$_{1-x}$Ir$_{x}$Te$_{2}$},    
    pdfauthor={M. Mandal, R. P. Singh},     
    pdfsubject={},   
    pdfcreator={},   
    pdfproducer={}, 
    pdfkeywords={} {} {}, 
    pdfnewwindow=true,      
    colorlinks=true,       
    linkcolor=blue, 
    citecolor=blue,        
    filecolor=magenta,      
    urlcolor=blue           
} 
\usepackage[normalem]{ulem}

\newcommand{\equref}[1]{Eq.~(\ref{#1})}

\newcommand{\figref}[1]{Fig.~\ref{#1}}

\newcommand{\tableref}[1]{Table~\ref{#1}}
\newcommand{\appref}[1]{Appendix~\ref{#1}}

\renewcommand{\approx}{\simeq}

\renewcommand{\vec}[1]{\boldsymbol{#1}}

\begin{document}

\title{\textrm{Superconductivity in doped Weyl semimetal Mo$_{0.9}$Ir$_{0.1}$Te$_{2}$ with broken inversion symmetry}}
\author{Manasi~Mandal}
\affiliation{Department of Physics, Indian Institute of Science Education and Research Bhopal, Bhopal, 462066, India}
\author{Chandan~Patra}
\affiliation{Department of Physics, Indian Institute of Science Education and Research Bhopal, Bhopal, 462066, India}
\author{Anshu~Kataria}
\affiliation{Department of Physics, Indian Institute of Science Education and Research Bhopal, Bhopal, 462066, India}
\author{Suvodeep~Paul}
\affiliation{Department of Physics, Indian Institute of Science Education and Research Bhopal, Bhopal, 462066, India}
\author{Surajit~Saha}
\affiliation{Department of Physics, Indian Institute of Science Education and Research Bhopal, Bhopal, 462066, India}
\author{R.~P.~Singh}
\email[]{rpsingh@iiserb.ac.in}
\affiliation{Department of Physics, Indian Institute of Science Education and Research Bhopal, Bhopal, 462066, India}
\date{\today}
\begin{abstract}

This work presents the emergence of superconductivity in Ir - doped Weyl semimetal T$_d$ - MoTe$_{2}$ with broken inversion symmetry. Chiral anomaly induced planar Hall effect and anisotropic magneto-resistance confirm the topological semimetallic nature of Mo$_{1-x}$Ir$_{x}$Te$_{2}$. Observation of weak anisotropic, moderately coupled type-II superconductivity in T$_d$ -Mo$_{1-x}$Ir$_{x}$Te$_{2}$ makes it a promising candidate for topological superconductor.


\end{abstract}


\maketitle

\section{INTRODUCTION}

Topological superconductor (TSC) is an exotic state of matter where the nontrivial topology of bulk promotes the emergence of Majorana bound states within the bulk superconducting gap \cite{TI2,TI3}. The realization of topological superconductors (TSCs) is a focus of research interest in condensed-matter physics due to their potential applications in fault-tolerant topological quantum computation. Theoretical studies have proposed that topological superconductivity can be induced at the interface of a topological material and a conventional superconductor \cite{TI4,TI5}. However, the possible complexity at the heterostructure fabrication extends the idea to search for superconductors with topological surface states. Recently, superconductivity in Weyl semimetal offers a new platform towards understanding topological superconductivity since the bulk nodal points have definite chirality and the Fermi surfaces are intrinsically topological \cite{TM1}. An interesting theoretical proposal suggests that the doped Weyl semimetals with four Weyl points are natural candidates to realize higher-order topological superconductors that exhibit a fully gapped bulk state, and the surface hosts robust gapless chiral hinge states. A finite-range attractive interaction in these materials favours p+ip pairing symmetry, and such a chiral pairing state is identified as a higher-order topological superconductor \cite{HOTI1}. 

MoTe$_{2}$ is a promising member of this family, having an abundant collection of structural variation with unusual attractive electronic properties, such as charge density wave (CDW) \cite{CDW}, an edge supercurrent \cite{edge_supercurrent} and surface superconductivity \cite{surface_superconductivity}. Noncentrosymmetric orthorhombic T$_{d}$-MoTe$_{2}$ was predicted to be type-II Weyl semimetal (WSM), which was further confirmed by experimental calculations like angle-resolved photoemission spectroscopy (ARPES) and scanning tunneling microscopy (STM) \cite{MoTe2_topology1,manasi_raman,MoTe2_topology7,MoTe2_STM}. T$_{d}$-phase MoTe$_{2}$ shows superconductivity at transition temperature T$_{C}$ = 0.10 K and superconducting transition temperature is sensitive to the chemical pressure \cite{MTS,MTSe1,MTSe2}. It is worthy to mention that S-doped MoTe$_{2}$ shows two band superconductivity with s$\pm$ pairing, and the quasi-particle interference patterns with band-structure calculations reveal the existence of Fermi arcs \cite{SMoTe2,SMoTe2_2}. Scanning tunnelling spectroscopy (STS) studies on the surface of doped MoTe$_{2}$ reveal a much larger superconducting gap that could be interpreted as topological nontrivial superconductivity based on the pairing of Fermi arc surface states \cite{SMoTe2}. Re doping in Mo-site in MoTe$_{2}$ enhances T$_{C}$ up to 4.1 K with new unconventional electronic properties like CDW, and strain-induced pseudo magnetic field and anomalous phonon behaviour \cite{manasi1,manasi2, jpcm,manasi_raman}. Also, the broken inversion symmetry in T$_{d}$ phase MoTe$_{2}$ can generate anti-symmetric spin-orbit coupling and lift the spin degeneracy. This allows an admixture of spin-singlet and spin-triplet superconducting ground states, which may lead to host exotic time-reversal symmetry breaking ground state \cite{ReZr,ReHf,Ti,ReTi2,La7Ir,La7Rh,La7Ni}. Therefore, it is essential to study superconductivity in MoTe$_{2}$ through chemical doping to identify the exotic topological states and determine the role of non-trivial topological surface states in the superconducting pairing mechanism.

In this paper, we report the emergence of superconductivity in T$_{d}$-Mo$_{1-x}$Ir$_{x}$Te$_{2}$ single crystal. It is characterised by Powder X-ray diffraction (XRD), energy-dispersive X-ray (EDAX) and  temperature dependent Raman measurements to confirm the phase purity and structural phase transition. Magnetization, AC transport, specific heat measurements were performed to probe superconductivity in Ir-doped MoTe$_{2}$ system. It is observed that Ir-substitution in Mo-site enhances the superconducting transition temperature up to 2.65(3) K. Planner Hall effect and anisotropic magneto-resistance confirm the Weyl semimetallic nature after the doping. Realisation of superconductivity with intact topological surface state suggests that Mo$_{1-x}$Ir$_{x}$Te$_{2}$ as a new candidate of topological superconductor.
    
\vspace{-0.3cm}
\begin{figure*}[htbp!]
\includegraphics[width=1.80\columnwidth]{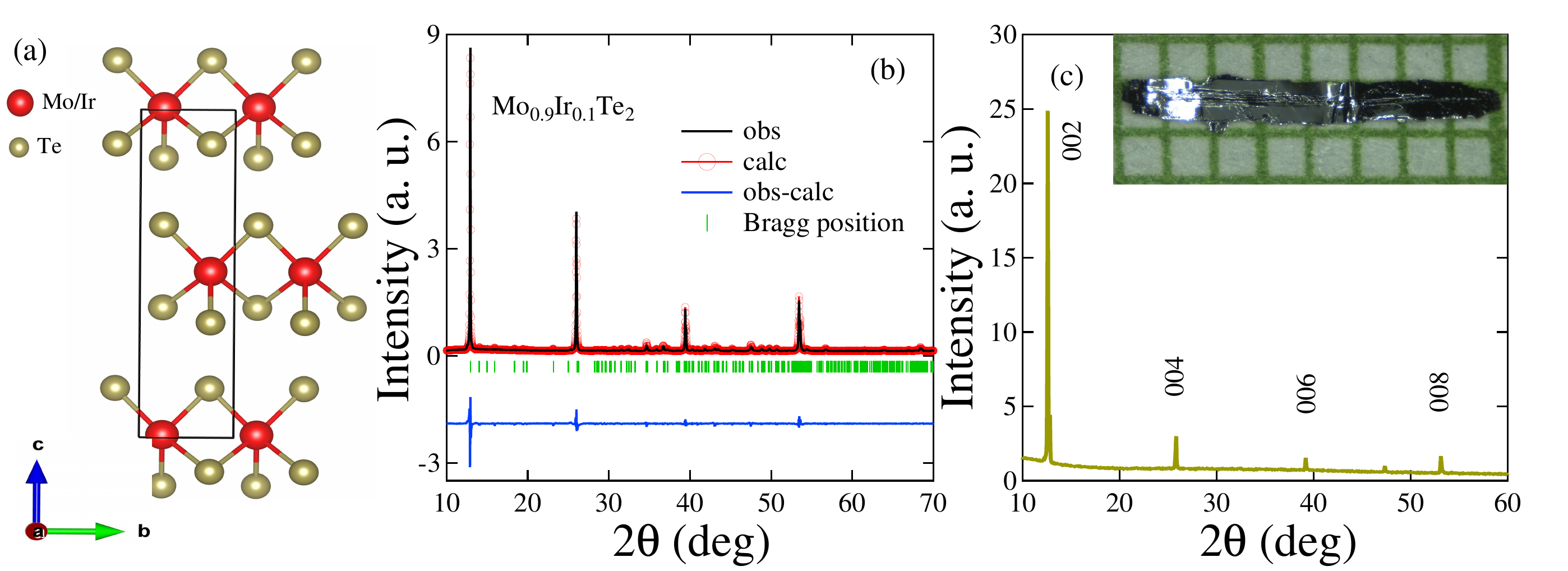}
\caption{\label{Fig1}(a) Room-temperature crystal structure of the Mo$_{0.9}$Ir$_{0.1}$Te$_{2}$ material (space group P21/m). (b) Rietveld refinement of powder XRD pattern for Mo$_{1-x}$Ir$_{x}$Te$_{2}$ (x = 0.1) recorded at room temperature. (c) Room temperature single crystal XRD shows the orientation of crystal growth along (00n) direction whereas inset shows the crystal image.}
\end{figure*} 

\section{EXPERIMENTAL DETAILS}

Pure phase polycrystalline samples of composition Mo$_{1-x}$Ir$_{x}$Te$_{2}$ (x = 0 and 0.1) were prepared from stoichiometric mixtures of Mo (99.9 \% pure), Ir (99.99 \% pure) and Te (99.99\% pure) by the standard solid-state reaction process \cite{manasi1}. We tried to dope Ir with a higher percentage, but 10\% was the highest possible concentration successfully achieved. Single crystals of Mo$_{1-x}$Ir$_{x}$Te$_{2}$ (x = 0.1) were grown by horizontal flux method with excess Te as flux \cite{flux_method}. Stoichiometric mixtures of Mo, Ir, and Te powders were sealed together in an evacuated quartz tube with charge and flux ratio 1:18. Crystallization was carried out at 1100$\degree$C for seven days in a box furnace, followed by the furnace switched off to avoid the hexagonal phase formation. Single crystals were trapped in the Te flux, which was taken out mechanically.

Powder x-ray diffraction (XRD) was carried out at room temperature on a PANalytical diffractometer equipped with Cu-K$_{\alpha}$ radiation ($\lambda$ = 1.54056 \text{\AA})for the characterization of crystal structure and phase purity. Rietveld refinement was performed using Full Prof Suite Software. Sample compositions were checked by a scanning electron microscope (SEM) equipped with an energy-dispersive X-ray (EDAX) spectrometer, which confirms Mo, Ir, and Te in the samples. The Raman spectra were measured in a Horiba JY LabRam HR Evolution Raman spectrometer and temperature-dependent Raman measurements were performed using a Linkam stage. To study the superconducting state, we measured DC and AC susceptibility by Superconducting Quantum Interference Device (SQUID MPMS, Quantum Design) at different fields. Specific heat measurements were performed by the two tau time-relaxation method in zero fields. Using the conventional four-probe technique with horizontal Rotator in the Physical Property Measurement System (PPMS, Quantum Design, Inc.), AC transport measurements were performed.

\section{RESULTS AND DISCUSSION}

\subsection{Sample characterization}
    
Polycrystalline sample was crushed in fine powder to perform XRD. The Rietveld refinement of XRD patterns confirms that the polycrystalline samples crystallized into single phase centrosymmetric monoclinic  CdI$_{2}$ type structure having space group P21/m as shown in \figref{Fig1} (b). XRD pattern of Mo$_{0.9}$Ir$_{0.1}$Te$_{2}$ single-crystal indicates the crystal orientation in (00n) direction (\figref{Fig1} (c)) and inset shows the crystal image. Raman spectra further confirms the phase purity (see \appref{Raman Study}). EDAX analysis shows the presence of Ir in the single crystal though the quantity of Ir is (4$\%$) lesser than the nominal one (see \appref{ExtractParameter}). 
   \begin{figure}[h!]
    \includegraphics[width=1.0\columnwidth]{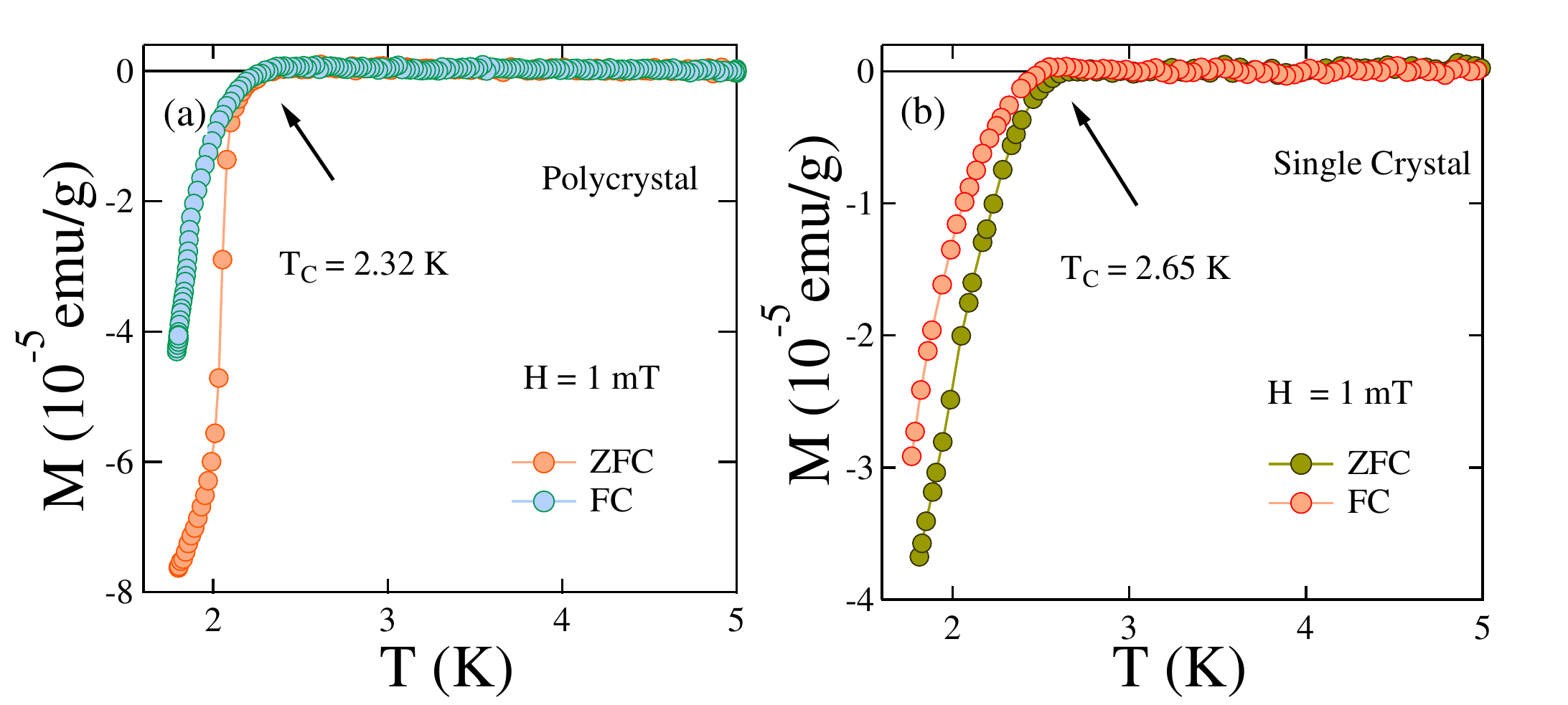}
    \caption {\label{Fig2} (a) and (b) Temperature dependence of the dc magnetic moment on polycrystal and single crystal sample at H = 1 mT.}
    \end{figure}

\subsection{Superconductivity}

Superconductivity in Mo$_{0.9}$Ir$_{0.1}$Te$_{2}$ polycrystal and the single crystal was further confirmed by magnetization measurements. Due to the small size of the crystals and less magnetic moments, a number of crystals were stacked together for magnetization measurements. The magnetic moment of the single crystal was checked in zero fields cooled (ZFC) and field cooled (FC) with 1 mT applied field. \figref{Fig2} exhibits a clear signature of type-II superconducting state with a diamagnetic signal. The polycrystalline sample shows superconducting transition temperature (T$_{C}$) at 2.32(2) K. In contrast, single-crystal yields T$_{C}$ at 2.65(3) K.
    
    \begin{figure}[h!]
    \includegraphics[width=1.0\columnwidth]{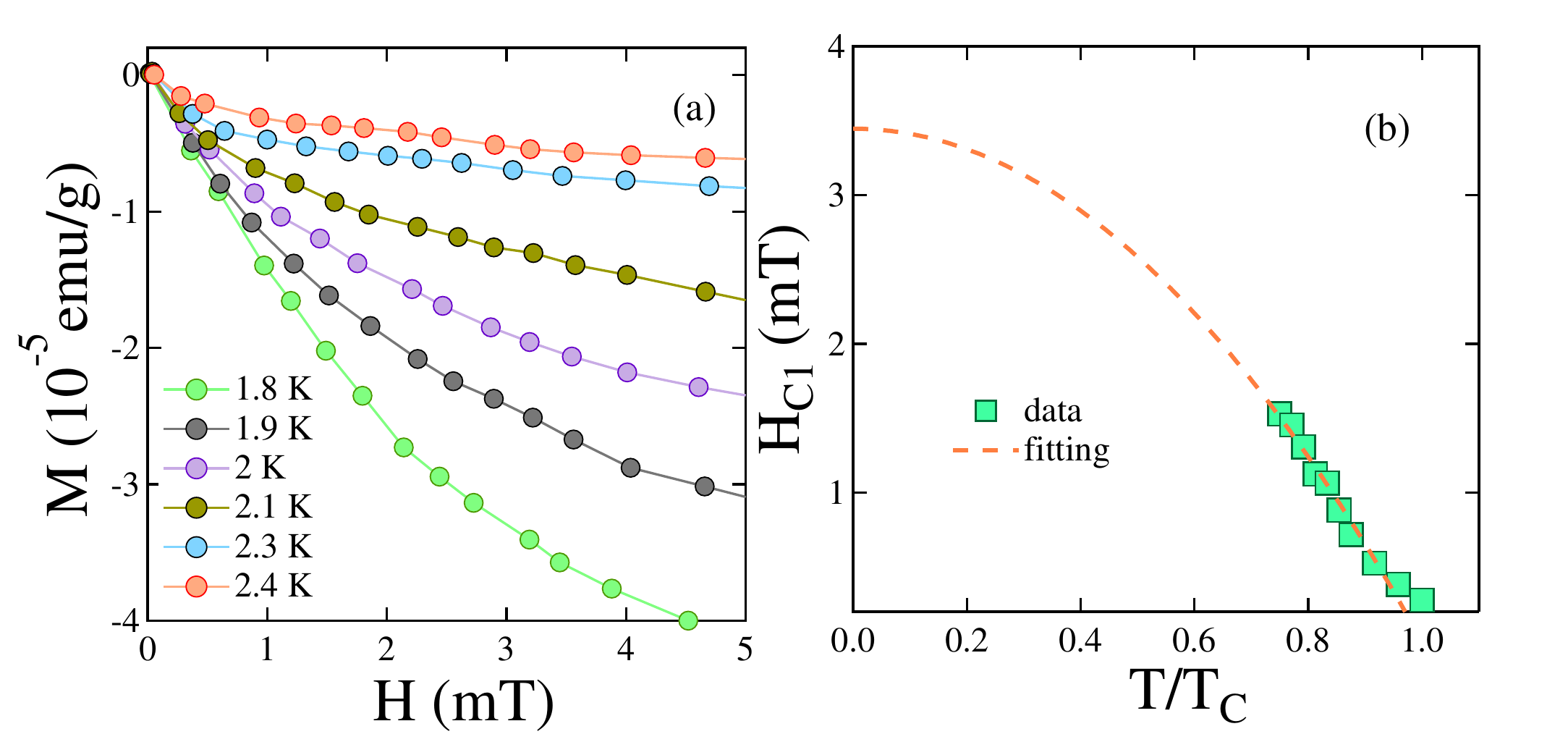}
    \caption{\label{Fig3}(a) The low-field magnetization curves for Mo$_{0.9}$Ir$_{0.1}$Te$_{2}$ single crystal at different temperature. (b) Temperature dependence of the lower critical field H$_{C1}$ was fitted using Ginzburg-Landau relation.}
    \end{figure}
    
    A sharp drop of electrical resistivity, $\rho$(T) further confirms the superconductivity at temperature, T$_{C}$ = 2.70(2) K (see \appref{AC Transport}). An anomaly with hysteresis in the $\rho$(T), which is associated with the first-order structural phase transition (T$_{S}$) from 1T${'}$ phase to T$_{d}$ phase is observed for Mo$_{0.9}$Ir$_{0.1}$Te$_{2}$ single crystal at 150 K, which is consistent with Raman study (see \appref{Raman Study}).\\
    
    To find the lower and upper critical fields, magnetic measurements were done on single crystal. We estimated the lower critical field H$_{C1}$(0) from the low field magnetization curves M(H) in the range of 0-5 mT took at different temperatures, as shown in (\figref{Fig3}). The lower critical field, H$_{C1}$(T), is determined from the first deviation from the linearity of the initial slope as the field is increased. H$_{C1}$(0) was estimated to be 3.45(4) mT by fitting the data in accordance with Ginzburg-Landau (GL) equation $H_{C1}(T)=H_{C1}(0)\left(1-\left(\frac{T}{T_{C}}\right)^{2}\right)$.
 
\begin{figure}[h!]
    \includegraphics[width=1.0\columnwidth]{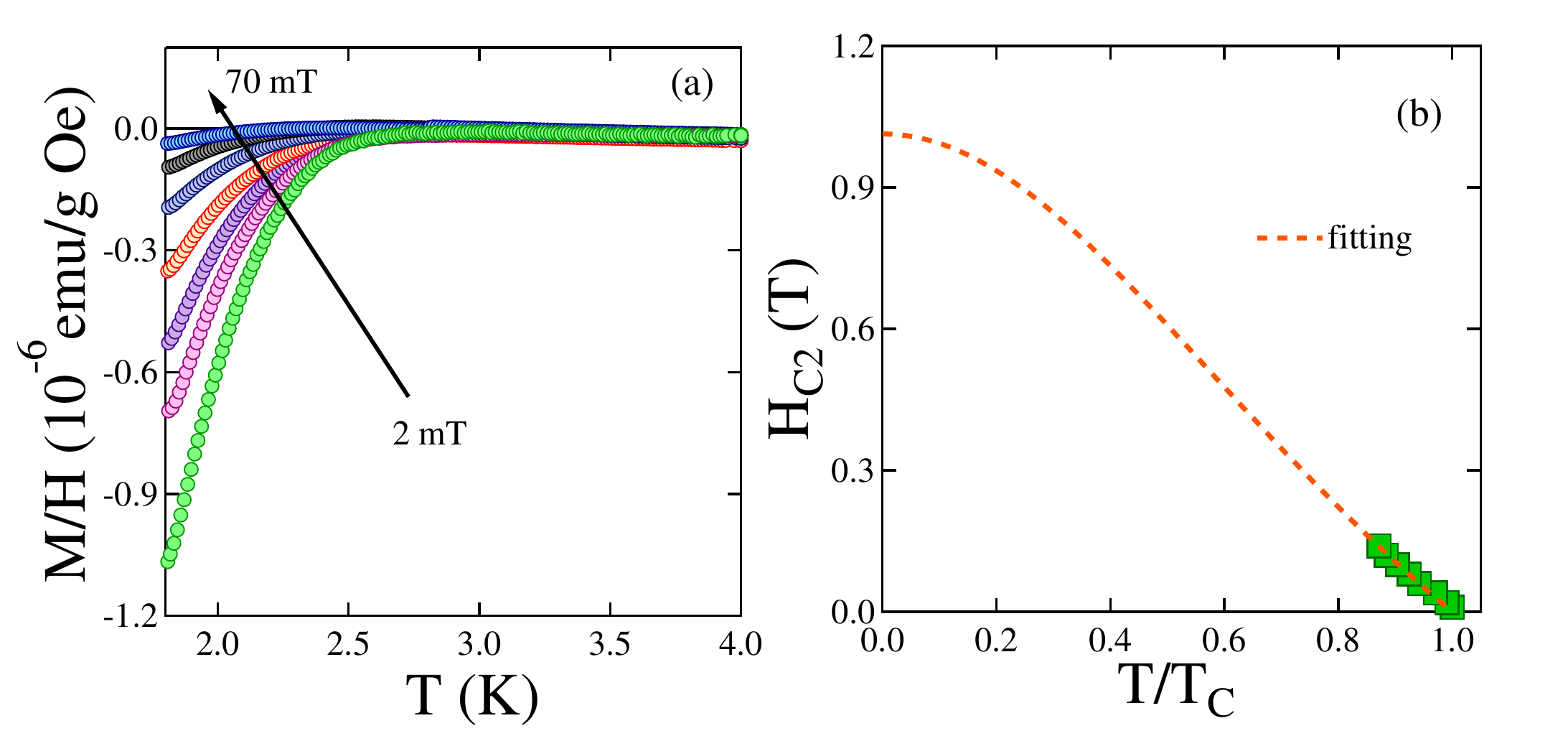}
    \caption {\label{Fig4}(a) The $\frac{M}{H}$ curves for various applied magnetic fields. (c) H$_{C2}$(0) was calculated by Ginzburg-Landau fits.}
    \end{figure}

    The temperature dependence of the magnetic moment under various magnetic fields was also measured to calculate the second-order transition field H$_{C2}$(0), as shown in \figref{Fig4} (a). Since, in the case of a type- II material, the magnetic field can penetrate the sample and reduce the gap function, T$_{C}$ shifts to lower temperature with the increment of the applied field. We estimated the value of the upper critical field by using Ginzburg-Landau (GL) formula $H_{C2}(T) = H_{C2}(0)\frac{(1-t^{2})}{(1+t^2)}$, where t = T/T$_{C}$. The estimated value of H$_{C2}$(0) is around 1.01(3) T as shown in \figref{Fig4}(b).

    The Ginzburg Landau coherence length $\xi_{GL}(0)$ is related to H$_{C2}(0)$  by the relation $H_{C2}(0)$ = $\frac{\Phi_{0}}{2\pi\xi_{GL}^{2}}$ where $\Phi_{0}$ is the magnetic flux quantum having value 2.07 $\times$10$^{-15}$ T m$^{2}$ \cite{Tinkham}. The another superconducting characteristics parameter Penetration depth $\lambda_{GL}$(0) associated with H$_{C1}$(0) by the relation \eqref{eqn8:PEN}:
    
    \begin{equation}
    H_{C1}(0) = \frac{\Phi_{0}}{4\pi\lambda_{GL}^2(0)}\left( ln \frac{\lambda_{GL}(0)}{\xi_{GL}(0)} + 0.497\right)
    \label{eqn8:PEN}
    \end{equation}
    
    The estimated values of $\xi_{GL}(0)$ and $\lambda_{GL}$(0) are 180.6 \text{\AA} and 2819 \text{\AA} respectively. GL parameter $k_{GL}$ = $\frac{\lambda_{GL}(0)}{\xi_{GL}(0)} $ = 21.2 ($\gg \frac{1}{\sqrt{2}}$), indicating that Mo$_{0.9}$Ir$_{0.1}$Te$_{2}$ is a type-II superconductor. Thermodynamic critical field H$_{C}$ is evaluated around 0.034(1) T using the relation $H_{C1}(0)H_{C2}(0)$ = $H_{C}^2lnk_{GL}$.
    Breaking of Cooper pair depends on the combine consequence of Pauli paramagnetic and orbital diamagnetic effect. The Pauli limiting field within the BCS theory is given by H$_{C2}^{p}$(0) = 1.83 T$_{C}$, which gives H$_{C2}^{p}$(0) = 4.85(1) T. In accordance to Werthamer-Helfand-Hohenberg (WHH) \cite{EH,NRW} expression
    the upper critical field in orbital limit, H$_{C2}^{orbital}$(0) can be calculated by the relation $H_{C2}^{orbital}(0) = -\alpha T_{C}\left. \frac{dH_{C2}(T)}{dT}\right|_{T = T_{C}}$. The obtained vale of H$_{C2}^{orbital}$(0) is 0.67(2) T using $\alpha$ = 0.69 for dirty limit superconductors and the initial slope $\left( \frac{-dH_{C2}(T)}{dT}\right)|_{T = T_{C}}$= 0.37 calculated from $H_{C2}$ -$\textit{T}$ phase diagram. The relative strengths of the orbital and Pauli limiting values of H$_{C2}$ is given by the Maki parameter $\alpha_{M} = \sqrt{2}H_{C2}^{orb}(0)/H_{C2}^{p}(0)$ \cite{maki}. We obtain $\alpha_{M}$ = 0.20 indicating that orbital effect is negligible in breaking of Cooper pairs.\\
   
   \begin{figure}[h!]
    \includegraphics[width=1.0\columnwidth]{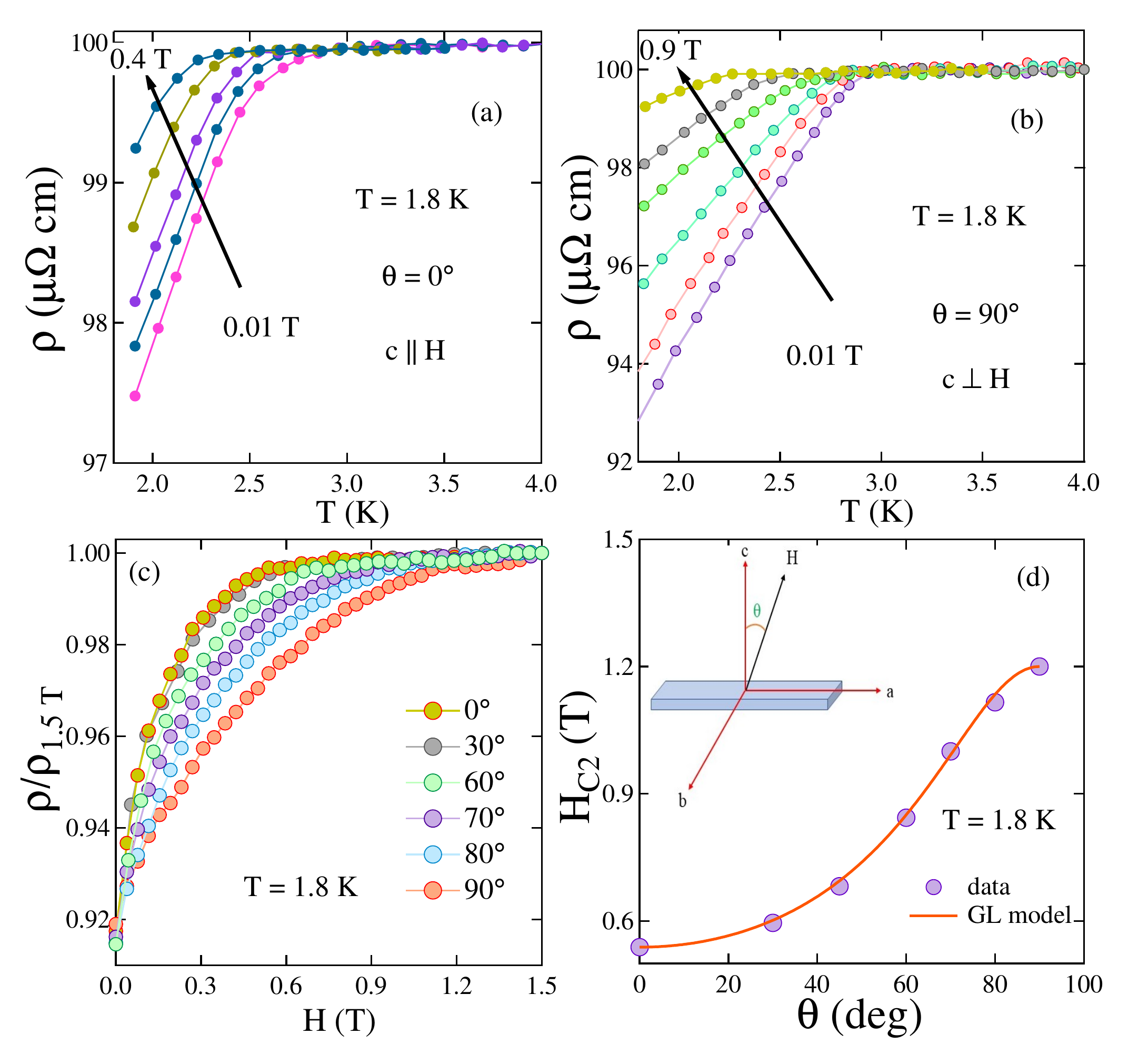}
    \caption {\label{Fig5} Temperature dependence of electrical resistivity $\rho$(T) at different field angle, $\theta$  = (a) 0$\degree$ and (b) 90$\degree$. (c) shows field dependence of electrical resistivity $\rho$(T) at different $\theta$ with fixed temperature, T = 1.8 K. (d) Angle dependent $\rho$(T) at T = 1.8 K whereas solid red line is 3D GL-fit (\equref{GL}).}
    \end{figure}
    
    We have further characterized the superconducting property by angle-dependent resistivity measurements. \figref{Fig5}(a) and (b) show the temperature dependence of resistivity for the magnetic fields H $\parallel$ c and H $\perp$ c, which clearly indicates anisotropy in the upper critical field, H$_{C2}$. \figref{Fig5}(c) shows the field dependence of the resistivity at different angles of H with c-axis at lowest possible temperature T = 1.8 K ($\approx$ 0.6 T$_{C}$). The angular dependence of H$_{C2}$($\theta$) determined from the onset value of the curve. In order to clarify the origin of the characteristic angular dependence of H$_{C2}$($\theta$), the experimental data were fitted with GL-model. According to the three dimensional (3D) anisotropic mass GL model H$_{C2}$($\theta$) can be described by the following relation \cite{CeIr3,NbSe2,NbSe2,NbSe2_2,NbSe2_3}:

    \begin{equation}
    \label{GL}
        \left(\frac{H_{C2}(T, \theta) cos\theta}{H_{C2}^{\perp}(T)}\right)^{2} + \left(\frac{H_{C2}(T, \theta) sin\theta}{H_{C2}^{\parallel}(T)}\right)^{2} = 1
    \end{equation}

\begin{figure}
    \includegraphics[width=1.0\columnwidth]{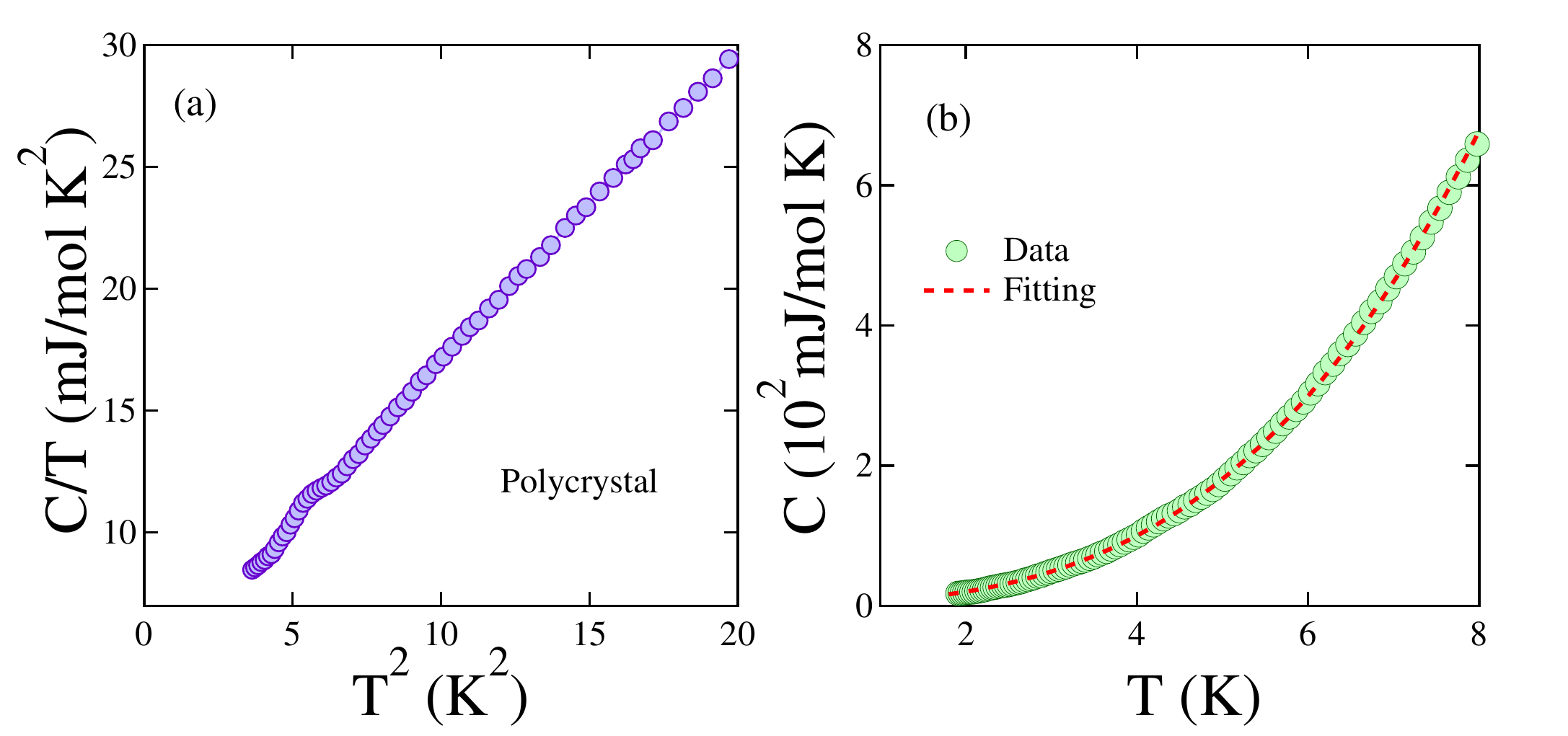}
    \caption{\label{Fig6} (a) C/T vs T$^{2}$ at 0 Oe field. (b) shows the data fitted with $\frac{C}{T}$ = $\gamma_{n}+\beta_{3}T^{2}+\beta_{5}T^{4}$.}
    \end{figure}

    The experimental data ware well fitted (red solid in \figref{Fig5}(d)) with the relation \equref{GL} with low anisotropy factor $\gamma$ = $\frac{H_{C2} (90\degree)}{H_{C2} (0\degree)}$ = 2.1. The nature of H$_{C2}$($\theta$) is in sharp contrast to pressure-induced two-dimensional superconductivity though the $\gamma$ value is comparable \cite{MoTe2}. Similar type of feature in H$_{C2}$($\theta$) has observed in 3D-anisotropic superconductor 2H-NbSe$_{2}$ \cite{NbSe2,NbSe2,NbSe2_2,NbSe2_3}, WTe$_{2}$ \cite{WTe2_hc2}, and gated MoS$_{2}$ \cite{MoS2}. Furthermore, the in-plane, $\xi_{GL}^{\parallel}$ and out-of-plane coherence lengths $\xi_{GL}^{\perp}$ at T = 1.8 K can be extracted from the H$_{C2}$ data, giving 247 and 166 \AA. The coherence length along the c direction ($\xi_{GL}^{\perp}$) is much larger than the lattice parameter along the same direction (13.7 \AA).\\
    Heat capacity measurements were performed at zero fields, where the superconducting transition is manifested by a jump in the heat capacity data at 2.4 K (\figref{Fig6}). The low temperature normal-state  specific heat data was analyzed in the temperature range 1.9 K $\le$ T $\le$ 8 K by using the relation $\frac{C}{T}$ = $\gamma_{n}+\beta_{3}T^{2}+\beta_{5}T^{4}$ where, $\gamma_{n}$ is Sommerfeld coefficient, $\beta_{3}$ is Debye constant and  $\beta_{5}$ is the anharmonic contribution to the specific heat. The extrapolation of normal state behavior below T$_{C}$, to the T $\to$ 0 limits, allows the determination of normal state coefficients. The linear fits to the C/T vs T$^2$ gave $\gamma_{n}$ = 5.05 $\pm$ 0.06 mJ mol$^{-1}$ K$^{-2}$, $\beta_{3}$ = 1.22 $\pm$ 0.01 mJ mol$^{-1}$ K$^{-4}$ and $\beta_{5}$ = (5 $\pm$ 0.1)*10$^{-7}$ $\mu$J mol$^{-1}$ K$^{-6}$ for Mo$_{0.9}$Ir$_{0.1}$Te$_{2}$. We calculated the Debye temperatures $\theta_{D}$ of the compounds using the formula $\theta_{D}$ = $\left(\frac{12\pi^{4}RN}{5\beta_{3}}\right)^{\frac{1}{3}}$ where  N (= 3) is the number of atoms per formula unit, R is the molar gas constant (= 8.314 J mol$^{-1}$ K$^{-1}$), which was based on the simple Debye model for the phonon contribution to the specific heat. The estimated $\theta_{D}$ value was 168.5 K. For non-interacting fermions, the Sommerfeld coefficient, $\gamma_{n}$ is proportional to the density of states $D_{C}(E_{F})$ at the Fermi level which was calculated to be 2.14 $\frac{states}{eV f.u}$ from the relation $\gamma_{n}$ = $\left(\frac{\pi^{2}k_{B}^{2}}{3}\right)D_{C}(E_{f})$, where k$_{B}$ $\approx$ 1.38 $\times$ 10$^{-23}$ J K$^{-1}$. The strength of the attractive interaction between electron and phonon can be estimated by the electron-phonon coupling constant, $\lambda_{e-ph}$ by McMillan equation \cite{Mc1} $\lambda_{e-ph} = \frac{1.04+\mu^{*}ln(\theta_{D}/1.45T_{C})}{(1-0.62\mu^{*})ln(\theta_{D}/1.45T_{C})-1.04 }$, where $\mu^{*}$ (= 0.13 for many superconductors) is the repulsive screened coulomb parameter. The calculated values of $\lambda_{e-ph}$ = 0.61 suggesting that Mo$_{0.9}$Ir$_{0.1}$Te$_{2}$ is a moderately coupled superconductor such as Mo$_{1-x}$Re$_{x}$Te$_{2}$ \cite{manasi1}, Mg$_{10}$Ir$_{19}$B$_{16}$ \cite{MgIrB}, and LaPtGe \cite{LPG}. The estimated parameters for Mo$_{0.9}$Ir$_{0.1}$Te$_{2}$ single crystal are comparable to Mo$_{1-x}$Re$_{x}$Te$_{2}$. We have summarized all the superconducting and normal state parameters in \tableref{tbl:parameters}. By substituting Ir for Mo in MoTe$_2$, facilitate the enhancement of the electron-phonon coupling and DOS at the Fermi energy similar to Re doped MoTe$_{2}$ compound \cite{manasi1}. So, we may expect unusual physical property in this system like Mo$_{1-x}$Re$_{x}$Te$_{2}$ compound. More investigation on a single crystal upto very low temperature is needed to explore the effect of Ir doping and the actual nature of the superconducting gap in this interesting system. To explore the topological nature in the Ir-doped MoTe$_{2}$ compound, we have done AC transport measurements.
    
    \begin{table}[h!]
    \caption{Normal and superconducting parameters of Mo$_{0.9}$Ir$_{0.1}$Te$_{2}$ }
    \setlength{\tabcolsep}{3pt}
    \begin{center}
    \label{tbl:parameters}
    \begin{tabular}[b]{l c  c c}\hline\hline
    Parameters& Unit& Mo$_{0.9}$Ir$_{0.1}$Te$_{2}$ & Mo$_{0.7}$Re$_{0.3}$Te$_{2}$\ \cite{manasi1}\\
    \hline
    \\[0.5ex]                                  
    T$_{C}^{res}$& K & 2.70(2) & 4.1  \\
    H$_{C1}(0)$& mT &  3.45(4) & 5\\ 
    H$_{C2}$(0)& T &  1.01(3) & 7.2\\
    H$_{C}(0)$& T & 0.034(1) & \\
    $\xi_{GL}$ & \text{\AA}&   180.6 & 72.2\\
    $\lambda_{GL}$ & \text{\AA}&    3819 & \\
    $k_{GL}$ & &  21.2 &\\
    $\gamma_{n}$ & mJ/mol K$^{2}$ & 5.05(6) &   9 \\
    $\beta_{3}$ & mJ/mol K$^{4}$ & 1.22(1) &   \\
    $\theta_{D}$ & K & 165.5 &   \\
    $\lambda_{e-ph}$ &  & 0.61 & 0.64  \\
    D$_{C}$(E$_{f}$) & states/eV f.u. &  2.14 &3.83   \\
    \\[0.5ex]
    \hline\hline
    \end{tabular}
    \par\medskip\footnotesize
    \end{center}
    \end{table}

\subsection{Topological Semimetallic Property}

    In transport experiments, WSMs usually distinctively attribute the chiral anomaly-induced negative longitudinal magnetoresistance (LMR) \cite{PHE1,PHE2,PHE3,PHE4}. This feature indicates the nonconservation of chiral charge around the Weyl nodes due to the coplanar electric and magnetic fields ($\vec{E}.\vec{B}$ = $\mathrm{E} \mathrm{B} \mathrm{cos \phi}$) \cite{PHE1,PHE2}. The experimental observation of negative LMR is very critical in type-II Weyl semimetal due to several extrinsic effects, such as the current jetting effect. Negative LMR has not been reported in the MoTe$_{2}$ system so far within our knowledge \cite{LMR,LMR_2}. Planar Hall effect (PHE) is also a practical tool to probe the chiral charge pumping effect that directly correlates the physics of topological nature and the transport properties. Experimentally PHE is observed in T$_d$-MoTe$_2$ \cite{MoTe2_topology2,MoTe2_topology4} as well as other Dirac semimetals such as WTe$_{2}$ \cite{WTe2,WTe2_2}, ZrTe$_5$ \cite{ZrTe5} and Cd$_3$As$_2$ \cite{CdAs,CdAs_2}.

     \begin{figure}[ht!]
    \includegraphics[width=1.0\columnwidth]{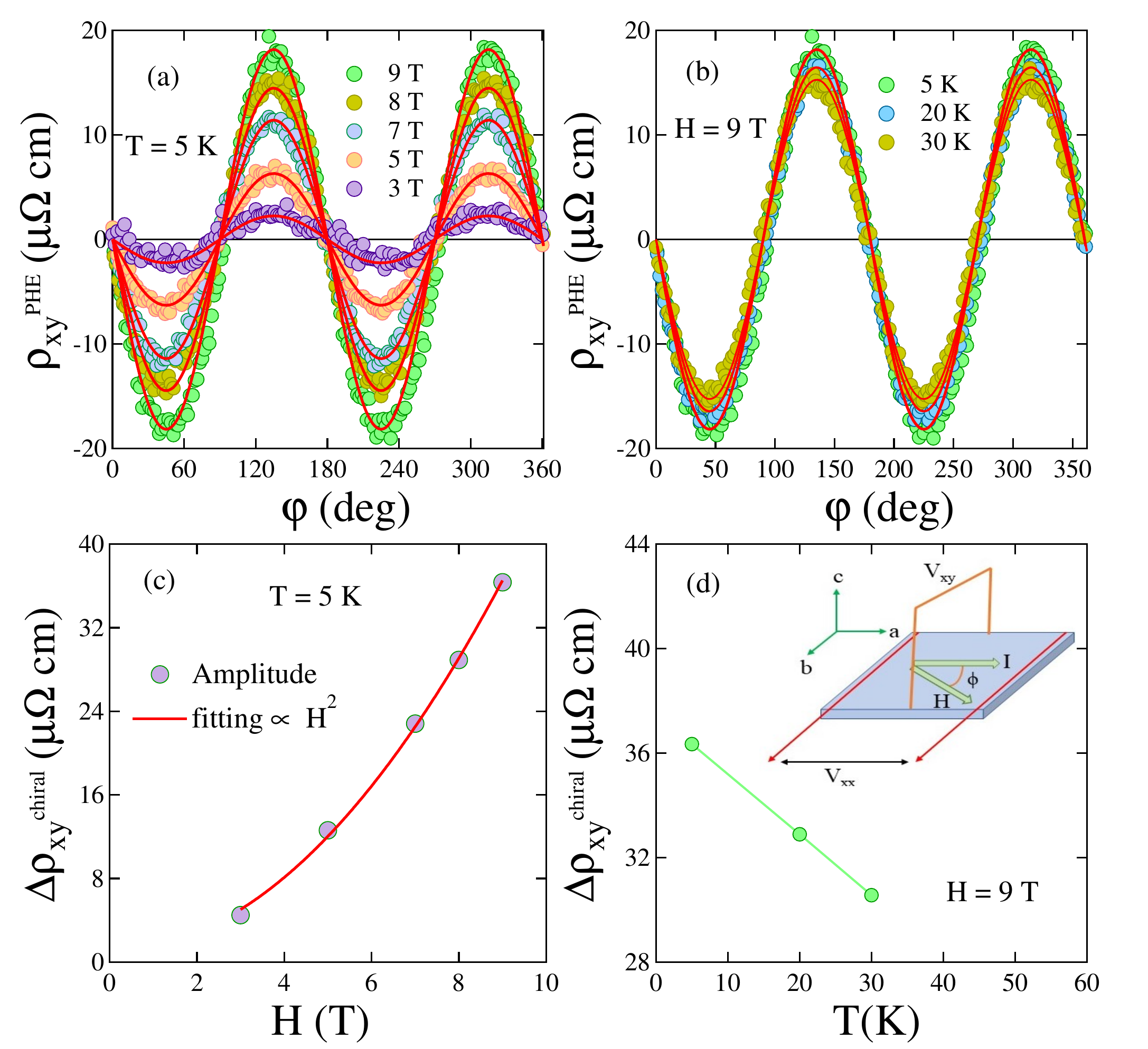}
    \caption {\label{Fig7} The extracted intrinsic planar Hall resistivity $\rho_{xy}^{PHE}$ (a) at T = 5 K under different fields (b) at H = 9 T under different temperatures. The variation of anisotropic resistivity, $\Delta\rho_{xy}^{PHE}$ with field and temperature is shown in (c) and (d).}
    \end{figure}
    
    We have measured planar Hall resistivity on Mo$_{0.9}$Ir$_{0.1}$Te$_{2}$ crystal where the applied field is rotated with angle $\phi$ to the current direction. A schematic of planar Hall measurements is shown in the inset of \figref{Fig7}. The obtained Hall resistivity has the total contribution of regular Hall and planar one. Normal Hall resistivity is antisymmetric under the opposite applied field, whereas the chiral-anomaly induced Hall resistivity is symmetric under the antisymmetric applied field \cite{PHE3,PHE4}. So planar hall resistivity can be extracted by average data of opposite applied field:

    \begin{equation}
    \rho_{xy}^{PHE} = [\rho_{xy}(B) + \rho_{xy}(-B)]/2  
    \end{equation}

    \figref{Fig7} (a) shows the angular dependence of planar Hall resistivity at T = 5 K under different applied magnetic fields. \figref{Fig7} (b) shows the same at H = 9 T under different temperatures. In WSMs, chiral anomaly induced planar Hall effect (PHE) can be expressed as \cite{PHE3,PHE4} 

    \begin{equation}
    \label{chiral_hall}
      \rho_{xy}^{PHE} = \Delta \rho_{xy}^{chiral} sin \phi cos \phi
    \end{equation}

    Where $\Delta \rho_{xy}^{chiral}$ is the chiral anomaly induced anisotropic resistivity that can be expressed as
$\Delta \rho_{xy}^{chiral} = \rho_{\perp} - \rho_{\parallel}$. Where $\rho_{\perp}$ and $\rho_{\parallel}$ are the resistivity corresponding to the magnetic field perpendicular to and along the direction of the current flow (I).

    \begin{figure}[h!]
    \includegraphics[width=1.0\columnwidth]{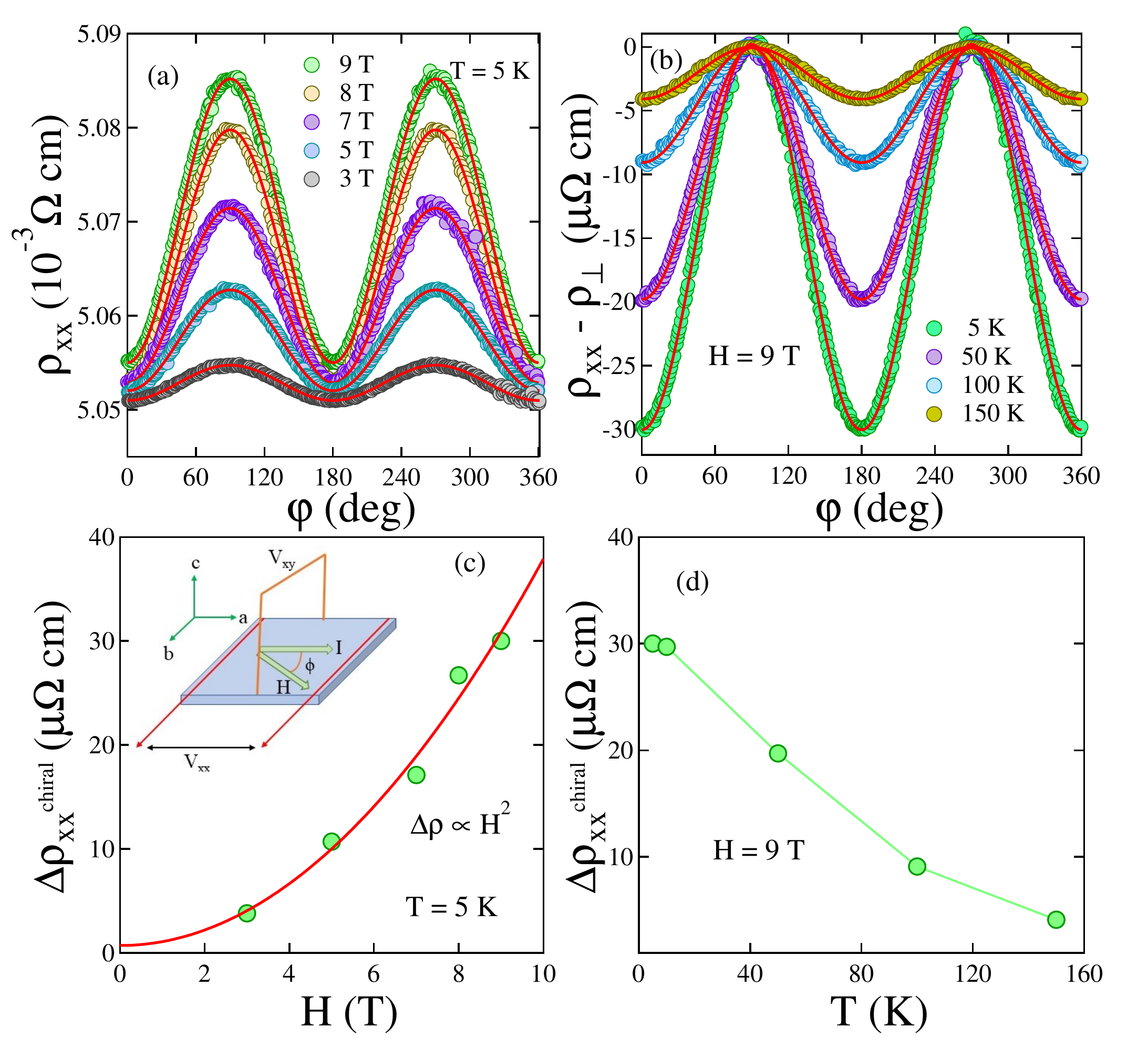}
    \caption {\label{Fig8} (a) Angle dependent $\rho_{xx}$ at T = 5 K under different fields whereas (b) shows the angle dependent ($\rho_{xx} - \rho_{\perp}$) at H = 9 T under different temperatures. The solid curve is a fit to equation. The variation of anisotropic resistivity, $\Delta\rho_{xx}$ with field and temperature extracted from fitting is shown in (c) and (d).}
    \end{figure}

    \figref{Fig7} shows that angular dependence of $\rho_{xy}^{PHE}$ has a period of $\pi$ with maximums appear at 45$\degree$ and 135$\degree$ similar to MoTe$_{2}$ and several other topological semimetals. We have fitted the data at T = 5 K with \equref{chiral_hall} and plotted the field dependence of $\Delta \rho_{xy}^{chiral}$ in \figref{Fig7}(c), which follows quadratic nature as expected for chiral anomaly induced PHE. Temperature dependence of $\Delta \rho_{xy}^{chiral}$ at H = 9 T is also shown in \figref{Fig7}(d) which decreases with increases temperature similar to the nature of DSMs and WSMs.
    
    Anisotropic magnetoresistivity (AMR), $\Delta\rho_{xx}$ can also be observed by the angular dependence ($\phi$) of longitudinal resistivity $\rho_{xx}$. In WSMs, chiral anomaly related AMR can be expressed as \cite{PHE3,PHE4}:

    \begin{equation}
    \label{AMR}
      \rho_{xx} = \rho_{\perp} - \Delta \rho_{xx}^{chiral} cos^{2} \phi  
    \end{equation}

    \figref{Fig8}(a) shows the the angular dependence ($\phi$) of $\rho_{xx}$ at T = 5 K under different applied fields. The data were fitted by the \equref{AMR}, and $\Delta\rho_{xx}^{chiral}$ was extracted for a different field. We have shown the field variation of $\Delta\rho_{xx}^{chiral}$ in \figref{Fig8}(c) that fitted well with quadratic equation ($\Delta\rho_{xx}^{chiral}$ $\propto$ H$^2$). The nature of the curve is consistent to \figref{Fig7}(c). The slight variation in value of  $\Delta\rho_{xx}^{chiral}$ and $\Delta\rho_{xy}^{chiral}$ be due to the sample dimensions used for the purpose of the measurements.

    \figref{Fig8}(b) shows the temperature dependence of $\rho_{xx} - \rho_{\perp}$ taken at various temperature at constant applied magnetic filed H = 9 T. The data were fitted with \equref{AMR} which yields temperature dependence of $\Delta\rho_{xx}^{chiral}$ as shown in \figref{Fig8}(d). Similar nature was observed in \figref{Fig8}(d) validating our measurements and analysis.
    
    It is observed that nontrivial Berry-curvature-induced PHE is prounced in Ir-doped T$_{d}$-MoTe$_{2}$ and still visible upto 150 K. The quadratic field dependence of $\Delta\rho_{xy}^{chiral}$ indicates the weak chiral anomaly induced coupling strength between electric and chiral charge whereas parent MoTe$_{2}$ shows intermediate coupling strength with high field at T $\leq$ 50 K \cite{MoTe2_topology4}. The change in coupling strength may be due to the small change in lattice constant or carrier concentration due to Ir-doping similar to the theoretically prediction \cite{theory_MoTe2,theory_MoTe2_2}.

\section{Conclusion}

    In summary, we have successfully prepared polycrystal and single crystal of Mo$_{1-x}$Ir$_{x}$Te$_{2}$ (x = 0 and 0.1) compound crystallized into a centrosymmetric monoclinic structure having space group P21/m. Magnetization measurements reveal that Mo$_{0.9}$Ir$_{0.1}$Te$_{2}$ is a type-II superconductor with a transition temperature of 2.65(3) K, which is 26 times enhanced with respect to parent MoTe$_{2}$. Resistivity measurements indicate that structural transition is decreased with Ir doping similar to the Re-doping effect in MoTe$_{2}$ compound. Raman study confirms the phase purity and 1T$^{\prime}$ to T$_{d}$ structural phase transition at T$_{S}$ $\approx $ 150 K. Angle-dependent higher critical field, H$_{C2} (\theta)$ follows 3D anisotropic GL-model with factor 2.1. The enhanced density of states, electron-phonon coupling, and other estimated values of superconducting and normal state parameters make Mo$_{0.9}$Ir$_{0.1}$Te$_{2}$ compound interesting to study further to explore unusual attractive electronic state like Re-doped MoTe$_{2}$ system. Above all, exotic chiral anomaly induced PHE is observed up to 150 K in Mo$_{1-x}$Ir$_{x}$Te$_{2}$ (x = 0.1). We can conclude this system as a possible topological superconductor with an exotic angle-dependent higher critical field. Further microscopic investigations like muon spectroscopy and scanning tunneling microscopy required to probe superconducting gap symmetry and ground state.

\section{Acknowledgments} R.~P.~S.\ acknowledges the Science and Engineering Research Board, Government of India for the Core Research Grant CRG/2019/001028. S.~S.\ acknowledges the Science and Engineering Research Board, Government of India for the Research Grant ECR/2016/001376. Authors thank Mr. Devesh Negi [CSIR fellow, File No.: 09/1020(0139)/2018-EMR-I] and Ms. Bommareddy Poojitha for their help with the Raman measurements. 


\appendix

\section{Characterization}\label{ExtractParameter}

    \begin{figure}[ht!]
    \includegraphics[width=1.0\columnwidth]{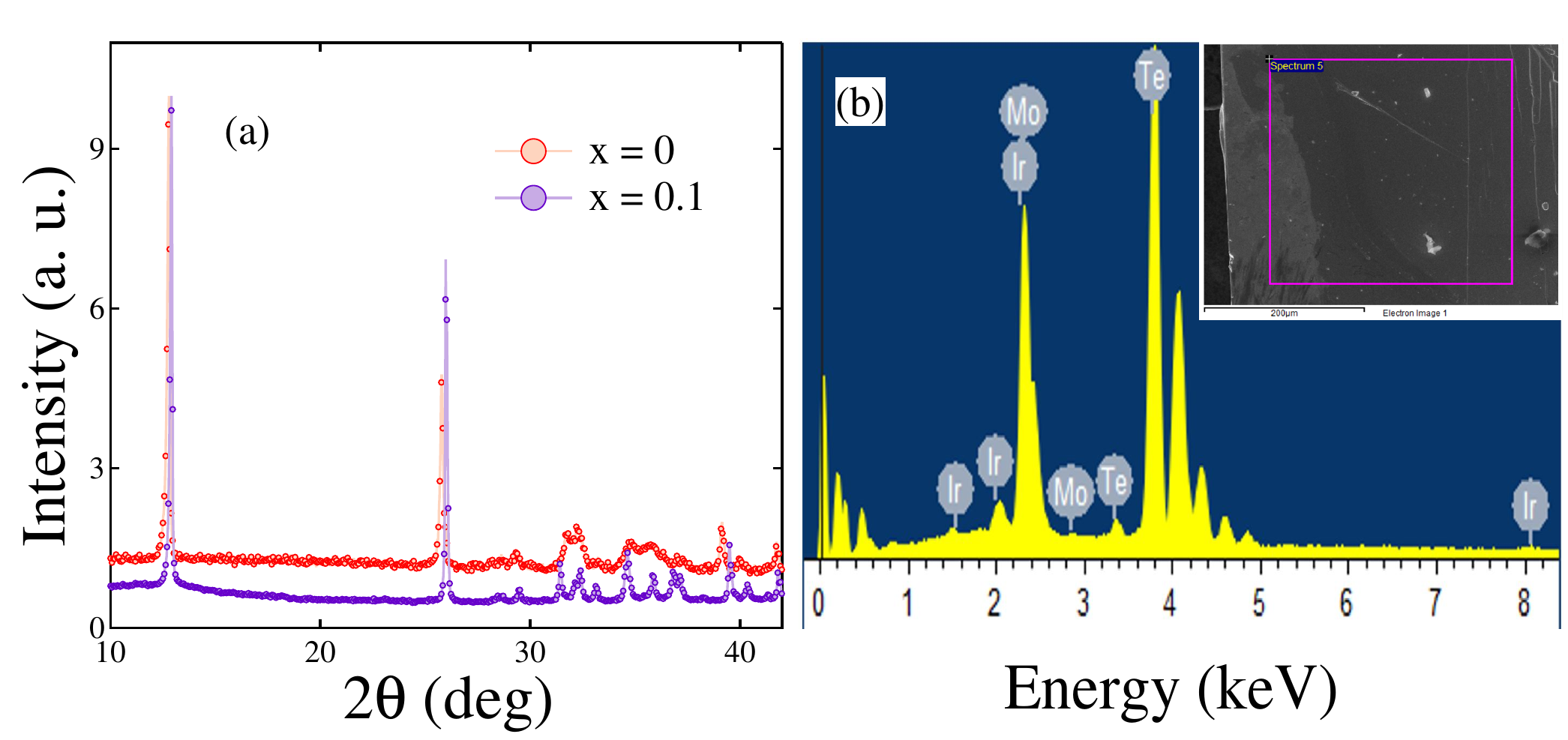}
    \caption {\label{Fig9} (a) Room temperature powder XRD pattern for x = 0 and 0.1 (b) EDX pattern for x = 0.1.}
    \end{figure}
    
    EDAX analysis is summarized in \tableref{tbl:EDS}.
    
    \begin{table}[h]
      \caption{Average elemental concentration (average of 10 points) obtained
    from the EDS measurements for Mo$_{1-x}$Ir$_{x}$Te$_{2}$ (x = 0.1) samples.}
      \label{tbl:EDS}
      \begin{tabular}{llc}
        \hline
        & Nominal composition & From EDS  \\
        \hline
        Polycrystal & Mo$_{0.9}$Ir$_{0.1}$Te$_{2}$  & Mo$_{0.9}$Ir$_{0.08}$Te$_{2}$ \\
       Single crystal & Mo$_{0.9}$Ir$_{0.1}$Te$_{2}$  & Mo$_{0.9}$Ir$_{0.04}$Te$_{2}$ \\
        \hline
      \end{tabular}
    \end{table}

\section{Raman Study}\label{Raman Study}
    
\begin{figure}[ht!]
    \includegraphics[width=1.0\columnwidth]{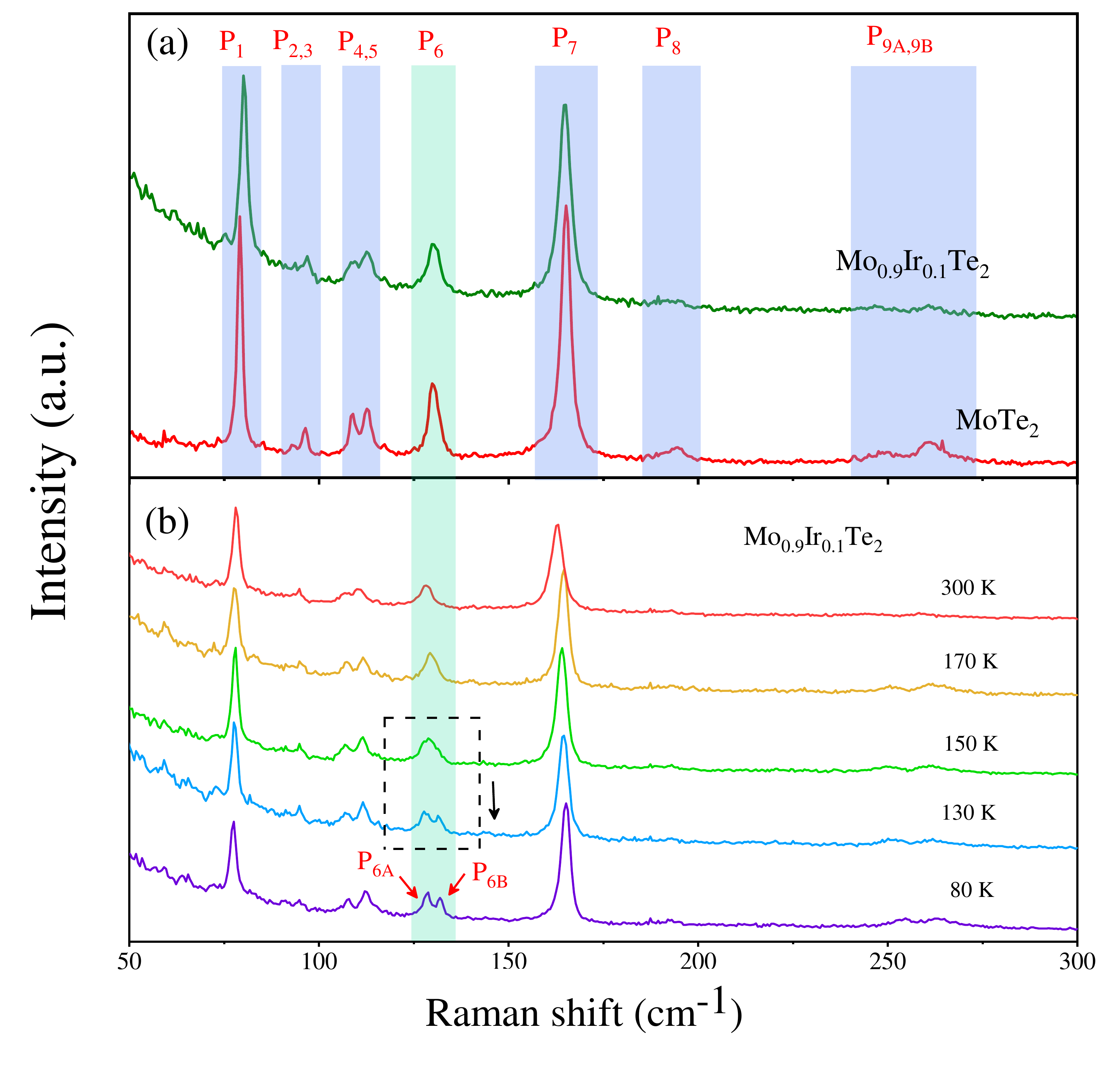}
    \caption {\label{Fig10} (a) A comparison of the room temperature Raman spectra of MoTe$_{2}$ and Mo$_{0.9}$Ir$_{0.1}$Te$_{2}$. (b) Raman spectra of Mo$_{0.9}$Ir$_{0.1}$Te$_{2}$ taken at different temperature revealing the splitting of the P$_{6}$ mode to P$_{6A}$ and P$_{6B}$ at low temperature.}
    \end{figure}

    Micro-Raman measurements are very effective in confirming the phase purity of investigated samples down to micron length scales. The micro-Raman measurements reported here were performed at a laser spot size of ~0.67 $\mu$m, thereby allowing the investigation of any impurity phase with a spatial resolution of $<1$ $\mu$m. \figref{Fig10}(a) compares the room-temperature Raman spectra of MoTe$_{2}$ and Mo$_{0.9}$Ir$_{0.1}$Te$_{2}$. The Raman modes observed in pure MoTe$_{2}$ are labelled as P$_{1}$-P$_{9}$, which correspond to out-of-plane vibrations with symmetries A$_{g}$ or B$_{g}$ \cite{raman1,raman2,raman3,raman4}. We also observe that the room-temperature Raman spectrum obtained for Mo$_{0.9}$Ir$_{0.1}$Te$_{2}$ shows similar modes as in for the pure phase, thus, confirming the absence of any impure phases in the investigated flakes. In addition to the confirmation of the phase purity, our temperature-dependent Raman measurements provide evidence of the topological phase transition in Mo$_{0.9}$Ir$_{0.1}$Te$_{2}$ at $\approx$150 K from a crystallographic point of view. \figref{Fig10}(b) shows the Raman spectra obtained for Mo$_{0.9}$Ir$_{0.1}$Te$_{2}$ at various temperatures. We clearly observe a splitting of the P6 mode into two bands (P$_{6A}$ and P$_{6B}$) below $\approx$150 K. A similar observation of mode-splitting was previously reported for pure \cite{manasi_raman,raman1,raman2,raman3,raman4} and Re-doped MoTe$_{2}$ \cite{manasi_raman}, attributed to the topological phase transition from the high-temperature 1T$'$ to low-temperature T$_{d}$ phase. Importantly, the topological phase transition is associated with a structural transformation characterized by a modification of the layer stacking order. The high-temperature phase (1T$'$) has a monoclinic structure possessing a center of inversion symmetry, while the low-temperature phase (T$_{d}$) is orthorhombic without the inversion symmetry. Therefore, the topological phase transition is associated with a breaking of inversion symmetry, which is also a fundamental requirement for the appearance of the Weyl semimetallic phase, that can be probed using Raman spectroscopy. The splitting of the P$_{6}$ mode is a result of the renormalization of phonons induced by the breaking of inversion symmetry at $\approx$ 150 K. The observation also matches well with the AC transport measurements (\appref{AC Transport}), which show a hysteretic change in $\rho$(T) around $\approx$150 K (\figref{Fig11}).
    
\section{AC Transport Measurements} \label{AC Transport}

    \begin{figure}[ht!]
    \includegraphics[width=1.0\columnwidth]{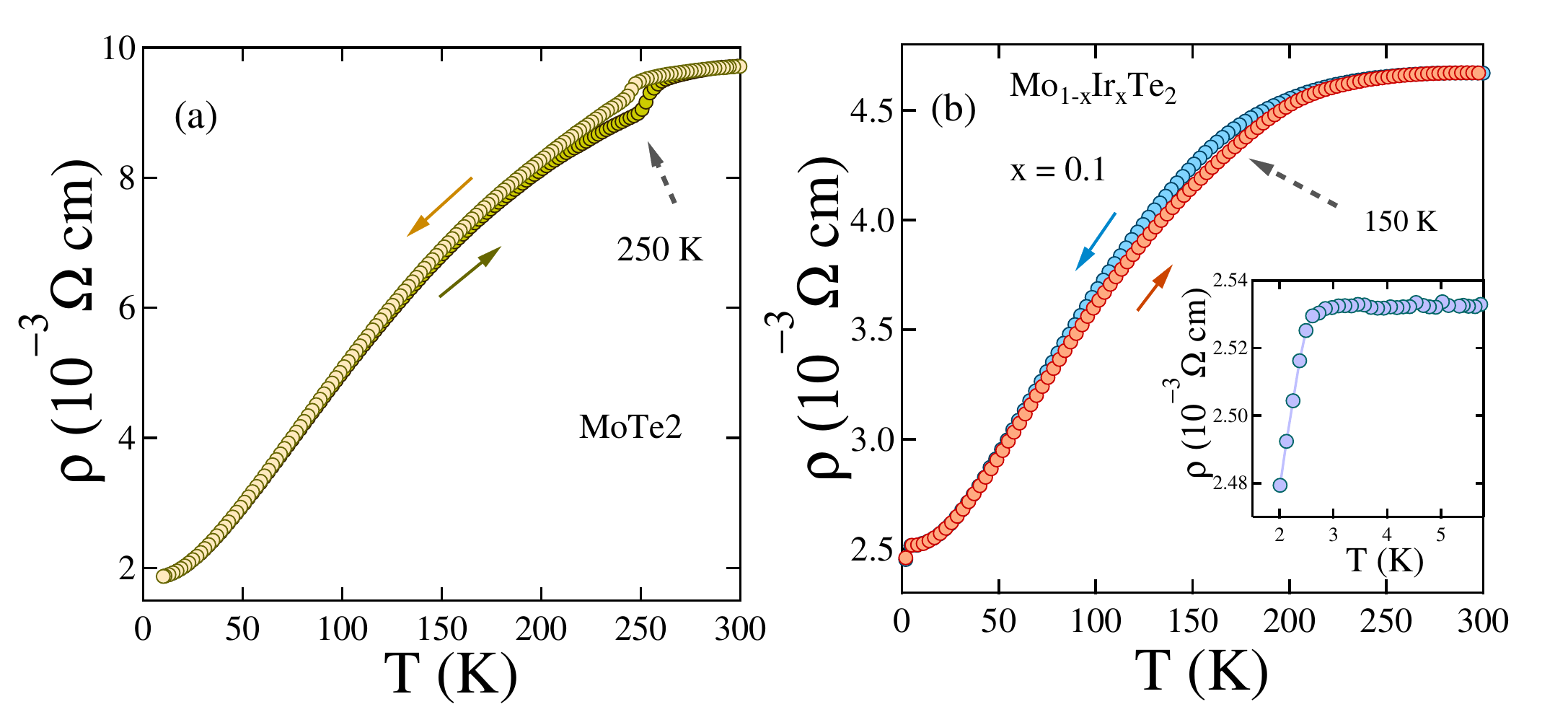}
    \caption {\label{Fig11} Normal state resistivity data $\rho$(T) over the temperature range 1.9 K $\le$ T $\le$ 300 K. A hysteresis in $\rho$(T) observed under the cooling and warming modes indicates polar structural transition temperature as shown by arrow symbol. Inset of (b) highlights the superconducting phase transition for x = 0.1.}
    \end{figure}

    Electrical resistivity, $\rho$(T) were measured on Mo$_{1-x}$ Ir$_{x}$Te$_{2}$ (x = 0 and 0.1) single crystal samples over the temperature range 1.9 K $\le$ T $\le$ 300 K in zero magnetic field in the cooling and warming modes. An anomaly with hysteresis in the $\rho$(T), which is associated with the first-order structural phase transition (T$_{S}$) from 1T${'}$ phase to T$_{d}$ phase is observed for both the samples (x = 0 and 0.1). The polar structural transition temperature, T$_{S}$, is found at around 240 K in parent MoTe$_{2}$ whereas the same has decreased to 150 K for Mo$_{0.9}$Ir$_{0.1}$Te$_{2}$, which is consistent with Raman study. The structural transition temperature is indicated by a arrow symbol in \figref{Fig11}. Similar type behaviour of resisitivity was observed in Re-doped MoTe$_{2}$ system \cite{manasi1}. A sharp drop of $\rho$(T) is observed in Mo$_{0.9}$Ir$_{0.1}$Te$_{2}$ sample at temperaute, T$_{C}$ = 2.70(2) K. This type of nature indicates the superconducting phase transition. 
    
    \begin{figure}[h!]
    \includegraphics[width=1.0\columnwidth]{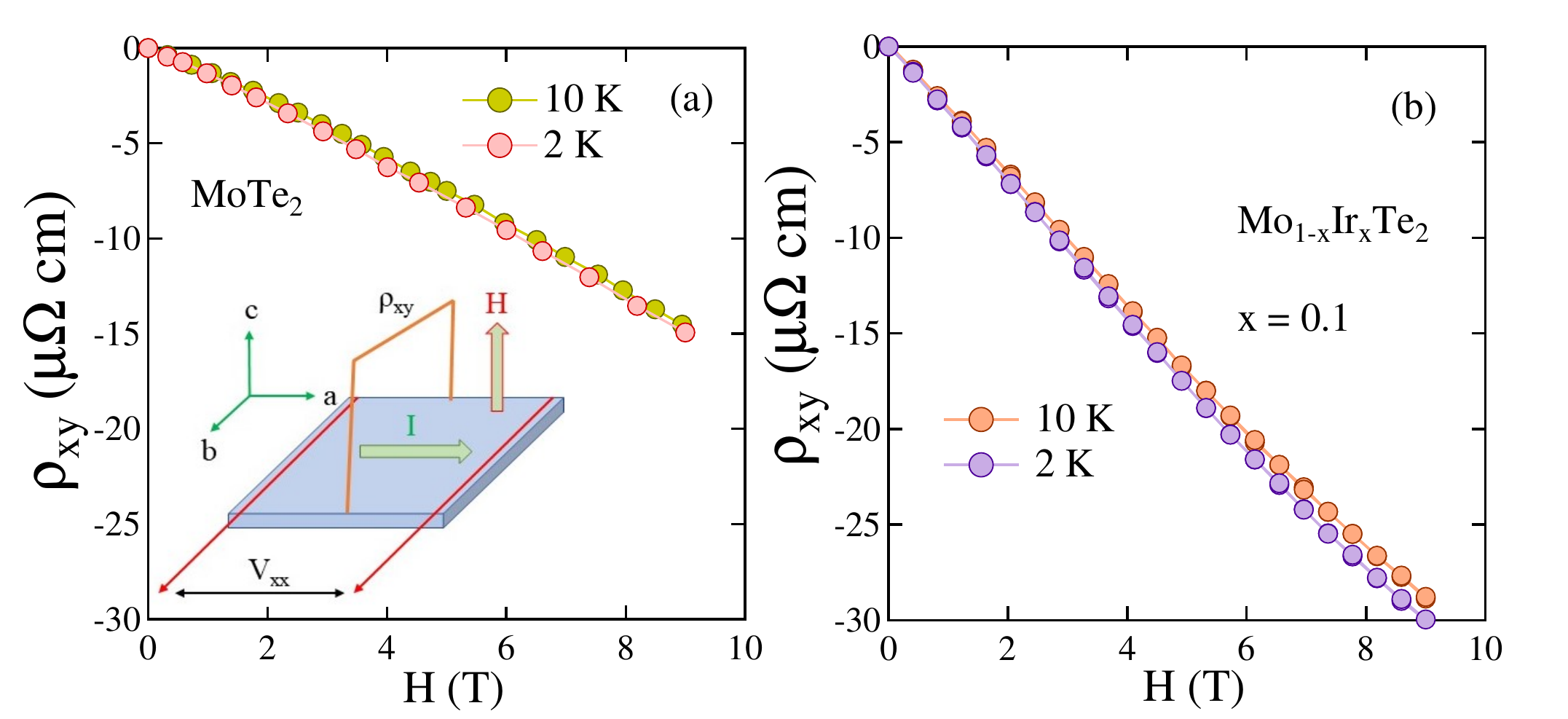}
    \caption {\label{Fig12} Regular Hall measurements performed on Mo$_{1-x}$ Ir$_{x}$Te$_{2}$ (x = 0 and 0.1) at 10 K and 2 K upto 9 T.}
    \end{figure}

    In order to find the nature of carrier we have performed regular Hall effect on the ab plane with the applied filed along the c-direction (Hall configuration is shown in \figref{Fig12}). The linear field dependence of hall resistivity shows negative slope indicating the dominating electron carrier in Mo$_{0.9}$Ir$_{0.1}$Te$_{2}$ system as shown in \figref{Fig12}. Ir doping may enhance the electron concentration in the system similar to Re doping in Mo-site \cite{manasi1}.

    \begin{figure}[h!]
    \includegraphics[width=1.0\columnwidth]{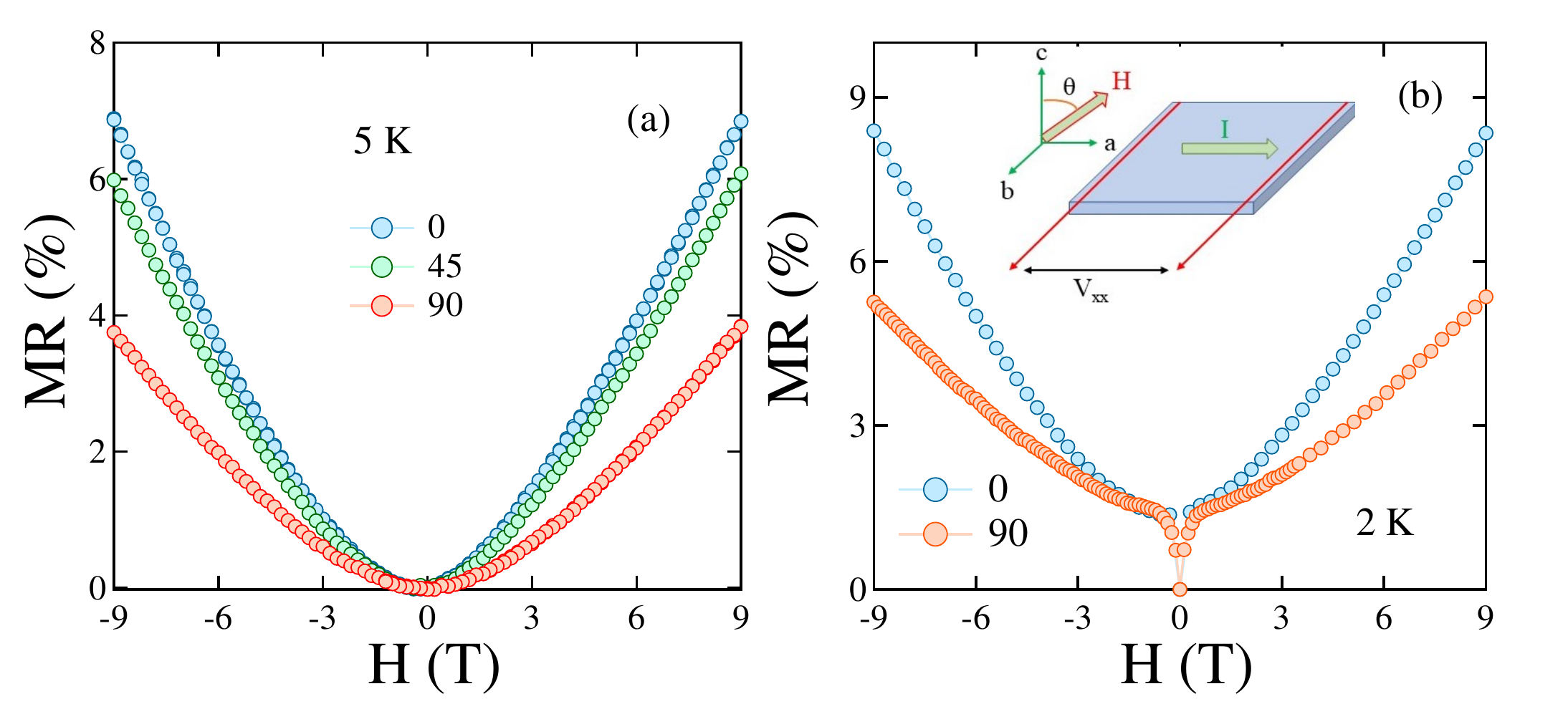}
    \caption {\label{Fig13} MR data taken at 5 K and 2 K on Mo$_{0.9}$Ir$_{0.1}$Te$_{2}$ crystal at different angles, $\theta$ between the direction of c-axis and applied magnetic field.}
    \end{figure}

    We have also measured magneto-resistance (MR) on Ir-doped crystal at 5 K and 2 K at different angles, $\theta$ between the c-axis and magnetic field direction, H. The the magnetic field dependence of the MR is defined as $\frac{\rho_{xx}(H)-\rho_{xx}(0)}{\rho_{xx}(0)}$. As shown in \figref{Fig13}(a) and (b), MR decreases with increasing angle $\theta$ from 0$\degree$ to 90$\degree$ at 5 K and 2 K, suggesting an anisotropy of transport behavior. MR data at T = 2 K indicates the superconducting nature at low field region. MR data suggests that magneto-resistance has been suppressed with Ir-doping in MoTe$_{2}$ system which may be due to the emergence of superconductivity \cite{MoTe2_topology4}.


\end{document}